# Amorphous Silicon Nanowires with Low Two-Photon Absorption, High Nonlinearity and Good Stability

C. Grillet [1,2], L. Carletti[2], C. Monat[2], P. Grosse[3], B. Ben Bakir[3], S. Menezo[3], J.M. Fedeli[3], and D. J. Moss[1]

[1] Institute of Photonics and Optical Sciences (IPOS) and CUDOS, School of Physics, University of Sydney, NSW 2006 Australia
[2] Université de Lyon, Institut des Nanotechnologies de Lyon (INL), Ecole Centrale de Lyon, 69131 Ecully, France
[3] CEA-Leti MINATEC Campus, 17 rue des Martyrs 38054 Grenoble Cedex 9, France

*Abstract* We demonstrate optically stable amorphous silicon nanowires with both high nonlinear figure of merit (FOM) of ~5 and high nonlinearity $Re(\gamma) = 1200 W^{-1} m^{-1}$.

## I. INTRODUCTION

Single crystal silicon-on insulator (SOI), compatible with computer chip technology (CMOS), has attracted huge interest over the past 10 years as a platform for nonlinear nanophotonic devices for all-optical signal processing [1], primarily because of its ability to achieve extremely high nonlinearities ($Re(\gamma) = \omega\, n_2 / c\, A_{eff}$, where $A_{eff}$ is the waveguide effective area) exceeding $300 W^{-1} m^{-1}$ [2] in nanowires and $6000\, W^{-1} m^{-1}$ in slow-light photonic crystal structure [3]. All-optical signal processing at bit rates of 160Gb/s [4] to over 1Tb/s [5], third harmonic generation [6], ultra-fast optical monitoring [7] and efficient correlated photon pair generation [8] have now been achieved. However, significant two-photon absorption (TPA) of crystalline silicon (c-Si) at telecom wavelengths is such that its nonlinear figure of merit (FOM = $n_2 / \beta_{TPA} \lambda$, where $\beta_{TPA}$ is the TPA coefficient and $n_2$ is the Kerr nonlinearity) is in the range of FOM= 0.3 to 0.5 [1] – much less than ideal for nonlinear optical applications. This has significantly limited the efficiency of these nonlinear devices - the largest parametric gain achieved in the telecom band in c-Si, for example, is only about 2dB [2].

The critical impact of a potentially low FOM was dramatically illustrated in recent experiments using new material platform with low TPA like chalcogenide [9], silicon nitride and hydex [10]. However, the low FOM of c-Si in the telecom band is a fundamental material property that cannot be improved.

It was recently suggested [11] that amorphous silicon (a-Si) could represent a promising alternative to crystalline silicon due to its expected lower nonlinear absorption resulting from the larger electronic bandgap of a-Si compared to c-Si. It can be deposited at low temperature and is compatible with current CMOS fabrication processes. Furthermore, unlike SOI it does not require epitaxial growth or wafer bonding to define the photonic layer and so it is also compatible with three-dimensional integration [12, 13]. Indeed, recent demonstrations [14, 15] have confirmed the possibility of increasing the FOM to as high as 2 at telecommunication wavelengths, allowing very high parametric gains of over +26dB over the C-band [14] to be achieved. However, to date a key drawback for this material has been a lack of stability, resulting in a dramatic degradation in performance over relatively short timescales (on the order of a few tens of minutes) [15]. Unless this problem can be solved, despite its very promising nonlinear performance, a-Si could well be in danger of becoming an academic curiosity. Here, we present amorphous silicon nanowires that exhibit very high optical nonlinearity together with simultaneous low nonlinear absorption and high optical stability [16]. This work paves the way for practical, highly efficient nonlinear photonic integrated circuits for ultrahigh speed optical signal processing [17].

## II. EXPERIMENT

The a:Si-H waveguides were fabricated in a 200mm CMOS pilot line at CEA-LETI, France. The a:Si-H film was deposited by plasma enhanced chemical vapor deposition (PECVD) at 350°C on 1.7µm oxide deposited on a bulk wafer. Serpentine waveguides with varying lengths (1.22 cm to 11 cm) were fabricated using DUV 193nm and dry etching. The fabricated waveguides are 220nm in thickness and ~ 500nm in width. The cross-section of the fabricated waveguide is shown in Fig. 1. A 500nm oxide was deposited to provide an upper cladding. The group velocity dispersion for the TE mode confined within a 500nm×220nm nanowire was calculated with FEMSIM, yielding an anomalous second-order dispersion parameter $\beta_2$=-$4.2\times10^{-25}$ $s^2$/m at $\lambda$=1550nm.

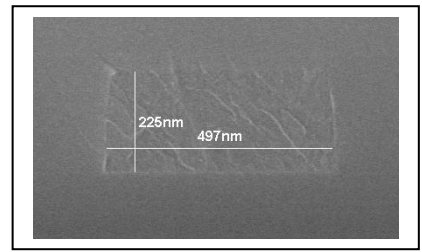

Figure 1. SEM cross-section of the a-Si nanophotonic wire embedded in silica

Figure 2 shows a schematic of the experimental setup used for the measurement of both the linear and nonlinear propagation characteristics of our a-Si nanowires. A mode-locked fiber laser with near transform limited ~1.8ps long pulses at a repetition rate of 20MHz at 1550nm was coupled into the TE mode of our a-Si nanowires via in-plane gratings. The fiber to waveguide coupling loss per coupler was ~ 10.6dB and 12.4dB per entry and exit, respectively, which was higher than expected due to the grating couplers not being optimized. The propagation loss of the TE mode was measured to be ~ 4.5 dB/cm, via a cut-back method on serpentine waveguides with lengths varying from 1.22 cm to 11 cm. In addition, a comparison with linear measurements performed on 1.3mm long straight nanowires yielded a loss contribution due to the bends of about 4dB, i.e. on the order of 0.04dB/ bend.

To determine the nonlinear parameters of our waveguides, we performed a series of self- phase modulation (SPM) measurements in a 1.22 cm long nanowire, with a coupled peak power up to ~3W. The output spectrum was then measured as a function of input power.

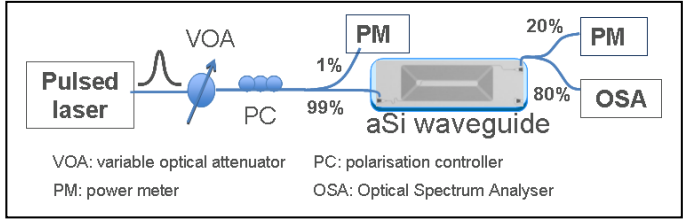

Figure 2. Experimental setup for observing SPM in a-Si:H waveguide.

### III. RESULTS AND DISCUSSION

Figure 3(top) shows the experimental contour plots of the output spectra as a function of coupled peak power ranging between 0.03W and 3W.

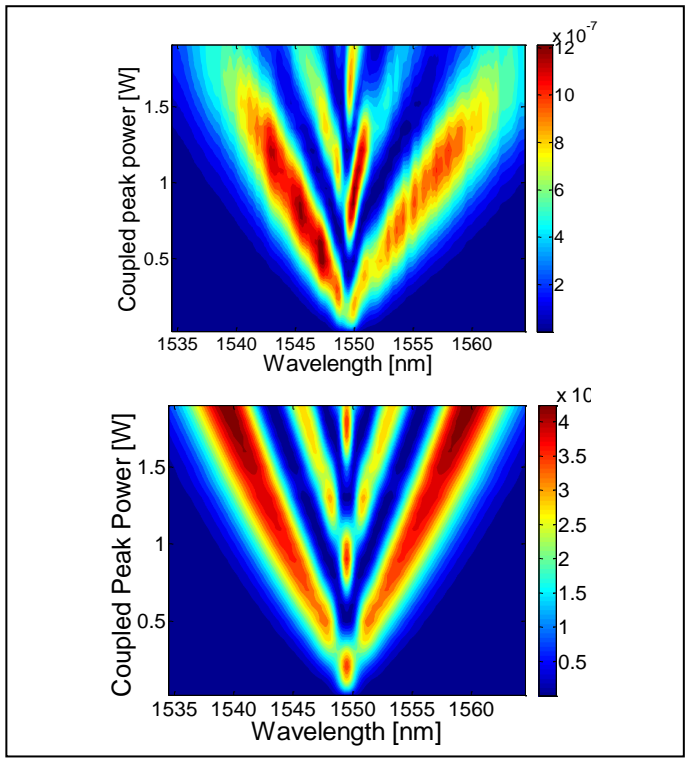

Figure 3. (top) Experimental and (bottom) theory 2D plots showing the spectral broadening of the output pulse spectra vs coupled peak power.

Strong spectral broadening is observed, which is the signature of self-phase modulation of the pulse propagating along the nonlinear nanowire. Note that for higher coupled powers (>1.8W), the spectral broadening of the pulse was limited by the spectral transfer function of the grating couplers, which have a 3dB bandwidth of around ~35nm centered near 1550nm.

In Figure 4, we plot the inverse of the waveguide transmission $T$ as a function of coupled peak power. The effect of the couplers can be seen at high powers on this curve as well. As the spectral broadening induced by SPM increases with the power, the lateral sidebands of the output pulse spectrum are attenuated by the transfer function of the couplers, causing a drop in the transmission (i.e. a more rapid increase in $1/T$). Due to this limitation, the $1/T$ curve is fit at low powers up to ~1.2W with the following linear equation valid in the presence of both linear propagation loss and TPA [16]:

$$\frac{1}{T} = \frac{P(0)}{P(L)} = 2\,\text{Im}(\gamma)L_{eff}e^{\alpha L}P(0) + e^{\alpha L} \quad (1)$$

where $P(0)$ and $P(L)$ are the optical peak power at the entrance (coupled peak power) and at the end of the waveguide, respectively, $\alpha$=179/m is the equivalent linear propagation loss including the bend loss contribution, $\text{Im}(\gamma)=\beta_{TPA}/(2A_{eff})$ is the imaginary part of the $\gamma$ nonlinear coefficient due to TPA, $L$ is the physical length and $L_{eff}$ the effective length reduced by the linear propagation loss through $L_{eff}=(1-e^{-\alpha L})/\alpha$. This allows us to extract $\text{Im}(\gamma)=18$/W/m±5%. Assuming a nonlinear modal area $A_{eff}$ of $0.07\mu m^2$, we infer a TPA coefficient $\beta_{TPA}$ equal to $0.25\times 10^{-11}$m/W±5%.

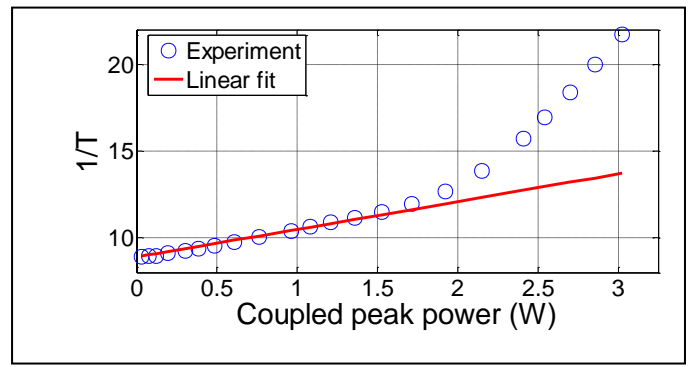

Figure 4. Inverse of the measured waveguide transmission versus coupled peak power (circles) along with a linear fit at low power.

Split-step-Fourier method simulations were then performed to solve the nonlinear Schrödinger equation governing the propagation of the picosecond optical pulse in the nonlinear waveguide, in the presence of second-order dispersion $\beta_2$ and TPA. The impact of dispersion was negligible, as expected for a $\beta_2$=-4.2×10$^{-25}$s$^2$/m, associated with a dispersion length exceeding 7m for the 1.8ps pulses, i.e. well over the physical length of the waveguide. Figure 3 (bottom) shows the output spectra resulting from these numerical simulations, showing a good agreement with the measurements, when taking the TPA contribution stated above and a nonlinear waveguide parameter Re($\gamma$)=1200/W/m, associated with a refractive index $n_2$ = 2.1 x $10^{-17}$ m$^2$/W±5%. Note that the low TPA of a:Si-H is also reflected in the absence of any blue shift in the output spectra.

The nonlinear index $n_2$ we obtain for this a-Si nanowire is about 4 times higher than c-Si. Together with the inferred TPA coefficient of 0.25 cm/GW, we estimate a FOM= 5±0.3 - an order of magnitude higher than c-Si and more than two times higher than any previous reported results in a:Si-H, combined with the high nonlinearity.

Most importantly, however, as mentioned a-Si has exhibited fundamental instability of its nonlinear characteristics when subjected to optical signals in the telecom band. In experiments with similar pulsed laser characteristics to ours, a significant degradation in the modulation instability (MI) sidebands was observed after only a few minutes. Even though this could be reversed under annealing, the post-annealed material still exhibited the same instability. Figure 6 shows the spectra out of our 1.22cm long waveguide recorded every two minutes when pumped with 2.25W of coupled peak power, corresponding to ~3GW/cm$^2$, over a timescale of 1hr.

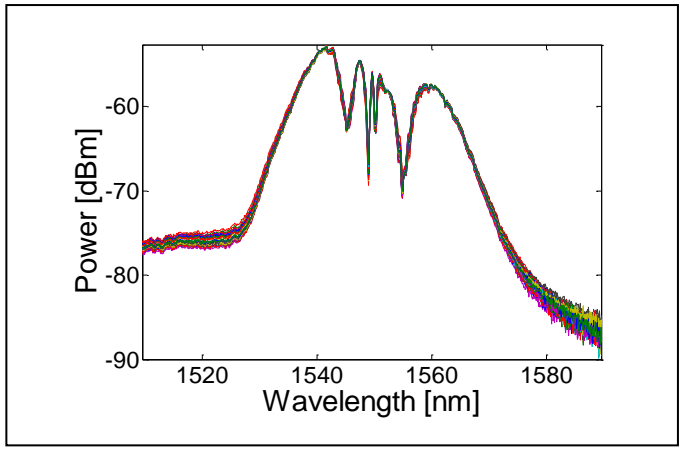

Figure 5. Output spectra as a function time for a coupled peak power of 1.5W. The spectra are recorded every 2min over 1 hour.

The spectra show negligible degradation in the SPM spectral pattern. Further, we observed no degradation in the parameters over the entire course of our experiments, which at times were conducted at peak powers of up to 3W (~4GW/ cm$^2$), primarily limited by the coupler insertion loss. We note that our material has also exhibited stable operation in the context of a hybrid integrated laser [14-15].

The decrease in TPA that has almost universally been observed in a:Si relative to c-Si in the telecom band is almost certainly due to an increase in the effective indirect bandgap in a:Si. Indirect, or phonon-assisted two-photon transitions have been shown [18-21] to account for TPA in c-Si in this wavelength range, which is not surprising since it is much less than half of the direct bandgap in bulk c-Si. However, the corresponding increase widely observed in $n_2$ in a:Si is quite surprising given that, in general, larger bandgaps yield smaller nonlinear response functions [22-25].

We believe that this issue could be fundamentally related to the relative contributions of indirect versus direct transitions to $n_2$ in bulk c:Si relative to a:Si. In bulk c=Si, full bandstructure calculations [22] of dispersion and anisotropy in the third order nonlinearity $\chi^{(3)}$ responsible for third harmonic generation [23-25] indicated that $\chi^{(3)}$ is dominated by direct transitions. For the component of $\chi^{(3)}$ responsible for $n_2$ and TPA in bulk c:Si, however, the situation is much less clear. Despite TPA in c:Si in the telecom band being dominated solely by indirect transitions, it is not at all obvious which transitions (direct or indirect) will dominate $n_2$ in this wavelength range since $n_2$ will contain contributions from *both* types of transitions, and generally speaking, contributions from indirect transitions tend to be one to two orders of magnitude weaker than those from direct transitions. If $n_2$ in this wavelength range were in fact dominated by direct transitions in c:Si, then it could easily be plausible that $n_2$ could increase in a:Si if the direct and indirect bandgaps moved in opposite diections relative to c-Si – ie., if a:Si had a smaller effective direct bandgap but a larger indirect bandgap relative to c:Si. We also note that in general momentum matrix elements play a key role in determining the magnitude of $n_2$ in any material, although it seems unlikely, a-priori, that these would vary by more than a factor of 2 or so in going from c:Si to a-Si. Although a very approximate estimate of the relative contributions of direct and indirect transitions to $n_2$ in c-Si seemed to indicate that indirect transitions dominate, this issue will only be properly resolved by a full bandstructure calculation of the dispersion in $n_2$ and TPA in c-Si and a:Si that includes both direct and indirect transitions, and this has not yet been reported.

In any event, the important point is that experimental evidence clearly shows that a:Si has the capacity to dramatically improve *both* the nonlinear FOM and $n_2$ in the telecom band, and with the combined demonstration of substantially increased optical stability reported here, this raises the prospect for a:Si to offer a very promising and practical platform for all-optical signal processing on a chip.

IV. CONCLUSION

We demonstrate a-Si:H nanowires with both high nonlinearity of Re($\gamma$) = 1200 W$^{-1}$m$^{-1}$ and nonlinear FOM of ~5. We observe no degradation in these parameters under systematic studies at 2.2W coupled peak power over timescales up to an hour, nor did we observe any degradation in the nonlinear parameters over the entire course of our experiments, at times up to 3W coupled peak power. These results represent the first report of simultaneously high FOM, nonlinearity and stability in a-Si:H and as such potentially pave the way for a-Si:H to offer a viable high performance nonlinear platform for all-optical devices in CMOS compatible integrated nanophotonic circuits, operating in the telecommunications window.

ACKNOWLEDGMENT

We acknowledge financial support of the European Union through the Marie Curie program (ALLOPTICS) and the Australian Research Council (ARC).

## REFERENCES

[1] J. Leuthold, C. Koos and W. Freude, "Nonlinear silicon photonics," Nature Photonics **4,** 535 (2010).

[2] M. Foster, A. C. Turner, J. E. Sharping, B. S. Schmidt, M. Lipson, and A. L. Gaeta, "Broad-band optical parametric gain on a silicon photonic chip," Nature **441**, 960-963 (2006).

[3] F. Li, M. Pelusi, D. X. Xu, A. Densmore, R. Ma, S. Janz, and D. J. Moss, "Error-free all-optical demultiplexing at 160Gb/s via FWM in asilicon nanowire," Opt. Express **18**, 3905-3910 (2010).

[4] H.Ji, M. Galili, H. Hu, M. Pu, L. K. Oxenlowe, K. Yvind, J. M. Hvam, and P. Jeppesen., "1.28-Tb/s Demultiplexing of an OTDM DPSK Data Signal Using a Silicon Waveguide", IEEE Photonics Technology Letters **22** 1762 (2010).

[5] C. Monat, B. Corcoran, D. Pudo, M. Ebnali-Heidari, C. Grillet, M. D. Pelusi, D. J. Moss, B. J. Eggleton, T. P. White, L. O'Faolain, and T. F. Krauss, "Slow light enhanced nonlinear optics in silicon photonic crystal waveguides," IEEE J. Sel. Top. Quantum Electron. **16**, 344–356 (2010)

[6] B. Corcoran, C. Monat, C. Grillet, D. J. Moss, B. J. Eggleton, T. P. White, L. O'Faolain, and T. F. Krauss, "Green light emission in silicon through slow-light enhanced third-harmoni generation in photonic crystal waveguides," Nature Photonics **3**, 206-210 (2009).

[7] B. Corcoran, C. Monat, et. al, "Optical signal processing on a silicon chip at 640Gb/s using slow-light," Opt. Express **18**, 7770-7781 (2010)

[8] C. Xiong, C. Monat, et. al, "Slow-light enhanced correlated photon pair generation in a silicon photonic crystal waveguide," Opt. Lett. **36**, 3413-3415 (2011)

[9] C. Monat, M. Spurny, C. Grillet, L. O'Faolain, T. F. Krauss, B. J. Eggleton, D. Bulla, S. Madden, and B. Luther-Davies, "Third-harmonic generation in slow-light chalcogenide glass photonic crystal waveguides," Opt. Lett. **36**, 2818-2820 (2011).

[10] L. Razzari, D. Duchesne, M. Ferrera, R. Morandotti, S. Chu, B. E. Little, and D. J. Moss, "CMOS-compatible integrated optical hyper-parametric oscillator," Nat. Photonics **4** (1), 41–45 (2010)

[11] K. Ikeda, Y. M. Shen, and Y. Fainman, "Enhanced optical nonlinearity in amorphous silicon and its application to waveguide devices," Opt.Express **15**, 17761-17771 (2007).

[12] B. Kuyken, et al., "On-chip parametric amplification with 26.5dB gain at telecommunication wavelengths using CMOS-compatible hydrogenated amorphous silicon waveguides," Opt. Lett. **36**, 552-554 (2011).

[13] B. Kuyken, et al.,"Nonlinear properties of and nonlinear processing in hydrogenated amorphous silicon waveguides," Opt. Express **19**, B146-B153 (2011).

[14] C.Sciancalepore, B. Ben Bakir, X. Letartre, J. Harduin, N. Olivier, C. Seassal, J-M. Fedeli, and P. Viktorovitch, "CMOS-Compatible Ultra-Compact 1.55- μm Emitting VCSELs Using Double Photonic Crystal Mirrors", IEEE Photonics Technology Letters **24**, 455 (2012).

[15] C.Sciancalepore, B. Ben Bakir, X. Letartre, J. Harduin, N. Olivier, C. Seassal, J-M. Fedeli, and P. Viktorovitch, "CMOS-Compatible Ultra-Compact 1.55- μm Emitting VCSELs Using Double Photonic Crystal Mirrors", IEEE Photonics Technology Letters **24**, 455 (2012).

[16] C.Grillet, C. Monat, L. Carletti, P. Grosse, B. Ben-Bakir, S. Menezo, and David J. Moss, "Amorphous Silicon Nanowires with Record High Nonlinearity, FOM, and Optical Stability", Optics Express **20** (20) 22609-22615 (2012). DOI: 10.1364/OE.20.022609.

[17] David J. Moss, R.Morandotti, A.Gaeta, M.Lipson, "New CMOS-compatible platforms based on silicon nitride and Hydex glass for nonlinear optics", Nature Photonics **7** (8) 597-607 (2013). DOI: 10.1038/nphoton.2013.183.

[18] M. Dinu, "Dispersion of Phonon-Assisted Nonresonant Third-Order Nonlinearities", IEEE J. of Quantum Electronics **39** 1498 (2003).

[19] J.L. Cheng, J. Rioux, and J. E. Sipe, "Full band structure calculation of two-photon indirect absorption in bulk silicon", Applied Physics Letters **98,** 131101 (2011).

[20] A.D. Bristow, N.Rotenberg, and H.M. van Driel, "Two-photon absorption and Kerr coefficients of silicon for 850–2200nm", Applied Physics Letters **90** 191104 (2007).

[21] M. Dinu, F. Quochi, and H. Garcia, "Third-order nonlinearities in silicon at telecom wavelengths", Applied Physics Letters **82**, 2954 (2003).

[22] D.J.Moss, J.Sipe, and H.van Driel, "Empirical tight binding calculation of dispersion in the 2nd order nonlinear optical constant of tetrahedral solids", Physical Review B **36,** 9708 (1987).

[23] D.J.Moss, E.Ghahramani, J.E.Sipe, and H.M. van Driel, "Band structure calculation of dispersion and anisotropy in χ$^{(3)}$ (3ω: ω,ω,ω) for third harmonic generation in Si, Ge, and GaAs", Physical Review B **41,** 1542 (1990).

[24] D.J.Moss, H.M.van Driel, and J.E.Sipe "Dispersion in the anisotropy for optical third harmonic generation in Si and Ge", Optics Letters **14,** 57 (1989).

[25] D.J.Moss, H.van Driel, and J.Sipe, "Third harmonic generation as a structural diagnostic of ion implanted amorphous and crystalline silicon", Applied Physics Letters **48,** 1150 (1986).
DOI: 10.1063/1.96453.